\documentclass[a4paper,11pt]{article}
\usepackage{jheppub}
\usepackage{lineno,color,slashed,changes}

\title{\boldmath $B^+\to K^+ \nu \bar{\nu}$ Excess and DM semi-annihilation}

\author[a,b,*]{Jongkuk Kim,}
\author[b,c]{and Pyungwon Ko}
\note[*]{Corresponding author.}

\affiliation[a]{
    Department of Physics,
    Chung-Ang University,\\
    Seoul 06974, Korea
}

\affiliation[b]{
	Excellence Cluster ORIGINS, \\
	Boltzmannstr. 2, D-85748 Garching, Germany}

\affiliation[c]{
    School of Physics, Korea Institute for Advanced Study, \\
    85 Hoegi-ro, Seoul 02455, Republic of Korea
}

\emailAdd{jongkukkim@cau.ac.kr}
\emailAdd{pko@kias.re.kr}

\abstract{
In 2023, Belle II collaboration announced the observarion of the 
$B^+ \to K^+ \nu\bar{\nu}$ decay channel for the first time.
This decay channel provides a clean signal with high precision in theoretical calculation.
However, we encounter $2.8\sigma$ deviation from the Standard Model (SM) prediction.
To resolve this excess, we study scalar dark matter (DM) model with local discrete $Z_3$ symmetry.
Assuming dark $U(1)_X \equiv U(1)_{L_\mu - L_\tau}$ symmetry, this $U(1)_{L_\mu - L_\tau}$ symmetry is spontaneously broken into local discrete $Z_3$ by non-zero vacuum expectation value of dark Higgs boson.
Considering dark Higgs mass is $2$GeV, we can explain the recent ${\rm Br} (B^+ \to K^+ \nu\bar{\nu})$ excess reported from Belle II collaboration and relic abundance at the same time.
}

\begin{document}
\maketitle
\flushbottom

\section{Introduction} \label{sec:intro}
In 2023, the Belle II collaboration reported the observation of the rare decay $B^+ \to K^+ \nu\bar{\nu}$, with a branching ratio exceeding the SM expectation by $\sim 2.8\sigma$ confidence level (C.L) \cite{Belle-II:2023esi}:
\begin{equation}
	{\cal B} ( B^+ \rightarrow K^+ \nu \bar{\nu} )_{\rm exp} = (2.3 \pm 0.7 ) \times 
	10^{-5}, 
\end{equation}
compared to the SM prediction,
\begin{equation}
	{\cal B} ( B^+ \rightarrow K^+ \nu \bar{\nu} )_{\rm SM} = (5.58\pm 0.37 ) \times 
	10^{-6}.
\end{equation}
This rare $B$ meson decay has long been recognized as a sensitive probe of physics 
beyond the SM due to the suppression in the Standard Model (SM) and the potential for contributions from new particles or interactions beyond the SM.
Although the current statistical significance is not yet sufficient to claim a new discovery, this mild deviation could be an early clue of new physics beyond the SM. 
A number of ideas have been proposed to interpret it in the context of various BSM's,from effective field theory approaches to concrete (UV-complete) models.
Several interpretations of this mild excess have been proposed in the literatures, including modifications of the Wilson coefficients ~\cite{Athron:2023hmz, Bause:2023mfe,
	Allwicher:2023xba,He:2023bnk,Hou:2024vyw}, 
three-body decay scenarios such as $B^+ \to K^+ \chi\bar{\chi}$ involving a new light particle $\chi$~\cite{Berezhnoy:2023rxx,Datta:2023iln,Altmannshofer:2023hkn,McKeen:2023uzo,
	Fridell:2023ssf,Cheung:2024oxh,Ho:2024cwk,Gabrielli:2024wys, Berezhnoy:2025nmb, Hu:2025zua}, 
as well as other BSM frameworks~\cite{Felkl:2023ayn,Wang:2023trd,He:2024iju,Bolton:2024egx,
	Rosauro-Alcaraz:2024mvx,Kim:2024tsm,Hati:2024ppg,Buras:2024ewl, Altmannshofer:2024kxb,Hu:2024mgf, Altmannshofer:2025eor,Calibbi:2025rpx,Lee:2025jky, He:2025jfc, Berezhnoy:2025tiw,Bolton:2025fsq,Aliev:2025hyp,Chen:2025npb, Ding:2025eqq, He:2025sao, DiLuzio:2025qkc, Shaw:2025ays, Berezhnoy:2025osn, Lee:2025kvf}. 
One tantalizing idea for the Belle II excess is to consider light dark sector for $B^+ \rightarrow K^+ +$ (missing energy).

Multiple observations from astrophysical and cosmological experiments strongly indicate the presence of DM.
The precise measurement of the cosmic microwave background (CMB) by the Planck Collaboration indicates that the total DM relic abundance is given by \cite{Planck:2018vyg}
\begin{align}
	\Omega_{\rm DM} h^2 &= 0.1200  \pm 0.0012.
\end{align}
Dark matter has been confirmed only through its gravitational effects, and its particle nature remains unknown.
This has prompted extensive theoretical studies focused on identifying viable DM candidates and the production mechanisms that could account for the observed relic abundance.
Among various DM models, the Weakly Interacting Massive Particle (WIMP) is a well motivated DM candidate.
The relic density of the thermal WIMP DM is determined by thermal freeze-out mechanism. 
The total WIMP DM annihilation cross section required to obtain the correct relic density is \cite{Steigman:2012nb}
\begin{align}
	\langle \sigma v \rangle_{\rm tot} \simeq \frac{1}{ (20 {\rm TeV})^2}
	\,.
\end{align}
Connecting these two seemingly unrelated observations, the Belle II excess and the dark matter relic density provides a good opportunity to explore some DM models.
In this work, we propose a framework that can simultaneously address the excess observed in $B^+ \to K^+ \nu\bar{\nu}$ at Belle II and account for the observed dark matter relic density through semi-annihilation in local $Z_3$ scalar DM models. 
By introducing a suitable mediator connecting the dark sector and SM neutrinos, our model provides an explanation of both phenomena at the same time. 

The paper is organized as follows.
We review local $Z_3$ scalar DM model in Sec.~\ref{model}.
We study DM phenomenology including relic density and direct detection bound in Sec.~\ref{DM:pheno}.
We explore the BelleII excess and possible solution in local $Z_3$ scalar DM model in Sec.~\ref{Belle2}.
Let us conclude in Sec.~\ref{con}.

\section{Local $Z_3$ scalar DM }  \label{model}
The dark ${\rm U(1)}_X\equiv {\rm U(1)}_{L_\mu - L_\tau}$ is anomaly-free, not requiring the introduction of additional chiral fermions~\cite{He:1990pn, He:1991qd}. 
In the minimal setup accommodating thermal WIMP DM, one introduces a dark Higgs $\Phi$, along with scalar dark matter candidate, $X$. 
Dark Higgs is related to the generation of $Z'$ boson mass.
We assume both the $\Phi$ and DM are singlets under the SM gauge group
${\rm SU}(3)_C \times {\rm SU}(2)_L \times {\rm U(1)}_Y$.
The charge assignments under 
the ${\rm U(1)}_{L_\mu - L_\tau}$ symmetry for these particles are specified as follows:
\begin{equation}
	Q_i ( \mu , \nu_\mu , \tau , \nu_\tau , X, \Phi ) = (1,1,-1,-1,Q_X, Q_\Phi ).
\end{equation}
The UV-complete Lagrangian is given by
\begin{eqnarray}
	\mathcal{L} &=& \mathcal{L}_{SM} -\frac{1}{4}Z'^{\mu\nu}Z'_{\mu\nu} - g_X Z'_\mu 
	\left( \bar{\ell}_\mu \gamma^\mu \ell_\mu - \bar{\ell}_\tau \gamma^\mu \ell_\tau +\bar{\mu}_R \gamma^\mu \mu_R - \bar{\tau}_R \gamma ^\mu \tau_R \right)\\
	& + & D_\mu \Phi^\dagger D^\mu \Phi - \lambda_\Phi \left( \Phi^\dagger \Phi - \frac{v_\Phi^2}{2} \right)^2 - \lambda_{\Phi H} \left( \Phi^\dagger \Phi - \frac{v_\Phi^2}{2} \right) \left( H^\dagger H  - \frac{v^2}{2} \right)  + \mathcal{L}_{\rm DM} ,    \nonumber
\end{eqnarray}
where $g_X$ is the ${\rm U(1)}_{L_\mu - L_\tau}$ gauge coupling constant, and $D_\mu = ( \partial_\mu + ig_X Q_i Z'_\mu )$ is the covariant derivative. 

After ${\rm U(1)}_{L_\mu - L_\tau}$ symmetry is spontaneously broken, the dark Higgs is described by
\[
\Phi (x) = \frac{1}{\sqrt{2}} \left( v_\Phi + \phi (x) \right) ,
\]
where $v_\Phi$ is the vacuum expectation value (VEV) of the dark Higgs field. 
The mass of  the dark photon $Z'$ is generated by the nonzero VEV of $\Phi$:
\begin{eqnarray}
	m_{Z'} &=& g_X \vert Q_\Phi \vert v_\Phi.
\end{eqnarray}

After acquiring nonzero VEVs in both the SM and $U(1)_{L_\mu - L_\tau }$ symmetry, two CP-even neutral scalar bosons mix with each other through the Higgs portal coupling, $\lambda_{\Phi H}$.
We define the mixing matrix $O$ connecting the interaction and mass eigenstates as follows:
\begin{equation}
	\left(
	\begin{array}{c}
		\phi \\ h
	\end{array}
	\right)
	= O 
	\left(
	\begin{array}{c}
		H_1 \\ H_2
	\end{array}
	\right)
	\equiv  
	\left(
	\begin{array}{cc}
		c_\theta & s_\theta \\
		-s_\theta & c_\theta 
	\end{array}
	\right)
	\left(
	\begin{array}{c}
		H_1 \\ H_2
	\end{array}
	\right),
\end{equation}
where $s_\theta (c_\theta) \equiv \sin\theta (\cos\theta)$, and $(\phi, h)$ and $(H_1, H_2)$ are the interaction and mass eigenstates with masses 
$m_{H_i}$ ($i=1,2$), respectively. 
The mixing angle $\theta$ is defined by
\begin{align}
	\tan 2\theta &= \frac{\lambda_{\Phi H} v_\Phi v_H}{\lambda_H v^2_H -\lambda_\Phi v^2_\Phi } ,
\end{align}
where $v_H \simeq 246$ GeV is the VEV of the SM Higgs.  
The mass matrix in the interaction basis $(\phi, h)$ can be expressed in terms of the 
physical parameters as follows:
\begin{equation}
	\left(
	\begin{array}{cc}
		2 \lambda_\Phi v_\Phi^2 & \lambda_{\Phi H} v_\Phi v_H \\ 
		\lambda_{\Phi H} v_\Phi v_H & 2 \lambda_H v_H^2
	\end{array}
	\right) =
	\left(
	\begin{array}{cc}
		m_{H_1}^2 {c_\theta^2} + m_{H_2}^2 {s_\theta^2} & (m_{H_2}^2 -m_{H_1}^2 ) c_\theta s_\theta \\ 
		(m_{H_2}^2 -m_{H_1}^2 ) c_\theta s_\theta & m_{H_1}^2 {s_\theta^2} + m_{H_2}^2 {c_\theta^2}
	\end{array}
	\right). \label{eq:mass_matrix}
\end{equation}
Now we can consider $m_{H_i}$ and $s_\theta$ as independent parameters.	
The mixing angle $\theta$ is constrained by the SM Higgs invisible decay data.
We focus on the small mixing, $\sin\theta \ll 1$.
Since our model describes a light dark Higgs scenario with $m_{H_1} < m_{H_2}$, we can consistently assume $|\theta| \ll \pi/4$.
Consequently, $H_1$ ($H_2$) is predominantly dark-Higgs-like (SM Higgs-like) particle, respectively.
We set $m_{H_2} = 125.25~\mathrm{GeV}$.

We assume that the kinetic mixing between the $Z'$ and the hypercharge gauge boson $B$ vanishes at a high scale.
However, a small mixing between the $Z'$ boson and the SM photon $A$ is generated radiatively.
After canonical basis of gauge fields by the transformation,
the interaction term is given by
\begin{align}
	\mathcal{L}_{\epsilon} &= - \epsilon_A Z'_\mu  J^\mu_{\text{em}},
\end{align}
where $J^\mu_{\rm em}$ denotes the electromagnetic current.
At the one-loop level, the mixing parameter $\epsilon_A$ is induced by the diagrams with $\mu$ and $\tau$ leptons running in the loop:
\begin{equation}
	\epsilon_A = -\frac{e g_X}{12\pi^2} \ln \left( \frac{m^2_\tau}{m^2_\mu} \right)
	\simeq -\frac{g_X}{70}.
\end{equation}

The long-standing discrepancy between the measured and SM predicted values of the muon anomalous magnetic moment, $a_\mu =(g-2)_\mu/2$, has long inspired various new physics scenarios.
However, the latest lattice-QCD–based SM prediction \cite{Aliberti:2025beg} suggests that the muong $(g-2)$ anomaly have been gone. 
The updated average including the recent Fermilab result \cite{Muong-2:2025xyk} gives
\begin{align}
	\Delta a_\mu = (39 \pm 64) \times 10^{-11}.
\end{align}
In this model, the contribution of the $Z'$ boson to $\Delta a_\mu$ at the one-loop level can be written as \cite{Baek:2001kca, Lynch:2001zr, Ma:2001md, Baek:2008nz, Altmannshofer:2014pba}
\begin{equation}
	\Delta a_\mu = \frac{g_X^2}{4\pi^2} \int_0^1 dx  \frac{m_\mu^2  x^2 (1-x)}{x^2 m_\mu^2 + (1-x) m_{Z'}^2}. \label{eq:a_mu_Zprime}
\end{equation}
In our analysis, we use the Fermilab measurement as an experimental constraint on the model parameter space, in particular on the gauge coupling $g_X$ and the dark photon mass $m_{Z'}$.
We adopt the current upper bounds on the gauge coupling $g_X$, for which $\Delta a_\mu \simeq 10 \times 10^{-10}$.

Constraints on the $(m_{Z'}, g_X)$ plane are taken into account from  measurements of $\Delta N_{\mathrm{eff}}$, and from the NA64 experiment.
The region $m_{Z'} \lesssim 10~\mathrm{MeV}$  is excluded by the measurements of $\Delta N_{\mathrm{eff}}$.
It is noteworthy that the Hubble tension can potentially be alleviated 
by the presence of a light $Z'$ contributing to a certain amount of dark radiation~\cite{Kamada:2015era, Escudero:2019gzq}.
After incorporating all experimental bounds, the viable parameter space consistent with $\Delta a_\mu$ in the case of a light $Z'$ is shown in Fig.~\ref{mZpVsgX}.
From now on we take $m_{Z'}=10$ MeV.

Let us consider complex scalar DM with charge assignments \(Q_\Phi = 3\) and \(Q_X = 1\), the \(U(1)_{L_\mu-L_\tau}\) symmetry is reduced to local discrete \(Z_3\) through the Krauss-Wilczek mechanism~\cite{Ko:2014loa,Ko:2014nha,Ko:2020gdg,Baek:2022ozm}. 
The Lagrangian for this is described by
\begin{align}
	\mathcal{L}_{\rm DM} =& \vert D_\mu X \vert^2 - m^2_X \vert X \vert^2 - \lambda_{HX} \vert X \vert^2 \left( \vert H \vert^2 - \frac{v^2_H}{2} \right) - \lambda_{\Phi X} \vert X \vert^2 \left( \vert \Phi \vert^2 
	- \frac{v^2_\Phi}{2} \right) + \lambda_3 \left(X^3\Phi^\dagger + \text{H.c.} \right) ,
\end{align}
where the last $\lambda_3$ term is a new gauge invariant operator specific for the charge assignments $Q_\Phi = 3 Q_X$. 
After $U(1)_{L_\mu-L_\tau} \rightarrow Z_3$, the scalar DM can be stable 
thanks to unbroken local $Z_3$ symmetry.
In our analysis, we mainly focus on the DM semi-annihilation processes, while keeping the $\lambda_{HX}$ and $\lambda_{\Phi X}$ terms in the Lagrangian for completeness.

The interaction between scalar DM and dark photon is given by
\begin{align}
	\mathcal{L}_{\rm DM} \supset ig_X \left( X^\dagger \partial_\mu X - X\partial_\mu X^\dagger \right)Z'^\mu. 
\end{align}
Through these interactions, DMs annihilate into a pair of leptons and neutrinos.

In the following, we examine the decay modes of the dark Higgs boson in detail. 
We investigate the possible decay channels of the dark Higgs boson. 
The dark Higgs can decay into a pair of gluons, photons, SM fermions, $Z'$ bosons, and dark matter particles.
The decay width of the dark Higgs into a pair of fermions is given by
\begin{eqnarray}
	\Gamma\left( H_1 \to f \bar{f}  \right) &=& 
	N_C s^2_\theta \frac{  m^2_f m_{H_1} }{8\pi v^2_H}  
	\left(1- \frac{4m^2_f}{m^2_{H_1}} \right)^{3/2} \Theta(m_{H_1} - 2m_f). 
\end{eqnarray}
The decay width of the dark Higgs into a pair of gluons is written as
\begin{eqnarray}
	\Gamma(H_1 \to gg) = s_\theta^2 \frac{\alpha_s^2 m_{H_1}^3}{72 \pi^3 v_H^2} 
	\left| \sum_q F_{1/2}(\tau_q) \right|^2,
\end{eqnarray}
where $\tau_q = \frac{4 m_q^2}{m_{H_1}^2}$.
Here, the sum runs over all SM quarks \(q\) in the loop, and the loop function \(F_{1/2}(\tau)\) is
\begin{eqnarray}
	F_{1/2}(\tau) = -2 \tau \left[ 1 + (1-\tau) f(\tau) \right],
\end{eqnarray}
with
\begin{eqnarray}
	f(\tau) = 
	\begin{cases}
		\arcsin^2 \frac{1}{\sqrt{\tau}}, & \tau \ge 1, \\
		-\frac{1}{4} \left[ \ln \frac{1+\sqrt{1-\tau}}{1-\sqrt{1-\tau}} - i \pi \right]^2, & \tau < 1.
	\end{cases}
\end{eqnarray}
The decay width into a pair of $Z'$ bosons is given by
\begin{eqnarray}
	\Gamma \left(H_1 \to Z'Z'  \right) &=& 
	\frac{g^2_X Q^2_\Phi c^2_\theta}{32\pi} 
	\frac{m^3_{H_1}}{m^2_{Z'}} 
	\left( 1 - \frac{4m^2_{Z'} }{m^2_{H_1}} + \frac{12m^4_{Z'} }{m^4_{H_1}}\right).
\end{eqnarray}
The decay width of the dark Higgs into a pair of DM particles is
\begin{align}
	\Gamma \left(H_1 \to X\bar{X} \right) 
	&= \frac{ \left( \lambda_{\Phi H}v_H s_\theta - \lambda_{\Phi X} v_\Phi c_\theta \right)^2 }{ 16\pi m_{H_1}} 
	\sqrt{1-\frac{4 m^2_X}{m^2_{H_1}} }.
\end{align}
Since we focus on the semi-annihilation process, we take $\lambda_{\Phi X}=\lambda_{HX}=0$. 
Therefore, the dark Higgs cannot decay into a dark matter pair at the tree level. 

To explain the Belle~II excess, the dark Higgs mass is required to be $\sim 2~{\rm GeV}$ \cite{Altmannshofer:2023hkn}. 
A detailed discussion of the Belle~II result will be presented in Sec.~\ref{BelleII}.
For $m_{H_1}=2~{\rm GeV}$, the dominant decay channel for the $H_1$ into the SM particles will be the gluon pair channel.
However, in our model, the branching ratio of the dark Higgs decay $H_1$ into 
a $Z'$ pair is almost unity.  And $Z'$ subsequently decays into a pair of neutrinos 
since the $Z'$ mass is lighter than the muon mass. 
The decay $Z' \to e^+ e^-$ is highly suppressed due to the small kinetic mixing, $\epsilon_A \sim - g_X/70 \sim \mathcal{O}(10^{-6})$. 
As a result of this tiny kinetic mixing, the decay width of the $Z'$ boson into an $e^+e^-$ pair is given by
\begin{align}
	\Gamma_{Z' \to e^+ e^-} = \frac{(e \epsilon_A)^2 m_{Z'}}{12\pi}
	\left( 1 + \frac{2 m_e^2}{m_{Z'}^2} \right)
	\sqrt{1 - \frac{4 m_e^2}{m_{Z'}^2}} \, .
\end{align}
This leads to the following branching ratio:
\begin{align}
	{\rm Br}(Z' \to e^+ e^-)
	= \frac{\Gamma_{Z' \to e^+ e^-}}
	{\Gamma_{Z' \to \nu_\mu \bar{\nu}_\mu} + \Gamma_{Z' \to \nu_\tau \bar{\nu}_\tau}}
	\simeq \left( \frac{e \epsilon_A }{g_X} \right)^{\!2}
	\simeq 2 \times 10^{-5}.
\end{align}

We now discuss the current bounds on the invisible decay of the SM Higgs boson. 
These bounds are primarily obtained from measurements at the LHC, by the ATLAS and CMS Collaborations, and constrain the branching ratio 
of the Higgs into invisible final states.  
Specifically, the ATLAS and CMS combined results~\cite{ATLAS:2023tkt, CMS:2023sdw} set an upper limit of \begin{equation}
	\mathrm{Br}(H_2 \to \mathrm{inv.}) \lesssim 0.11 \quad \text{at 95\% C.L.}
\end{equation}
Future collider experiments are expected to further improve these limits. 
For example, the High-Luminosity LHC (HL-LHC)~\cite{Cepeda:2019klc},  
the Compact Linear Collider (CLIC)~\cite{CLICdp:2018cto}, the Future Circular Collider in the $e^+e^-$ mode (FCC-$ee$)~\cite{Blondel:2021ema}, and 
the International Linear Collider (ILC)~\cite{Potter:2022shg} 
are projected to reach sensitivities down to
\begin{equation}
	\mathrm{Br}(H_2 \to \mathrm{inv.}) \sim 0.001.
\end{equation}
These constraints are crucial for testing models in which the Higgs boson can decay into dark matter or other light dark sector particles.

\begin{table}[tbh]
\centering
\begin{tabular}{|c|c|c|}
\hline
Experiments & ${\rm Br}(H_2 \to {\rm inv.})$ & Refs \\
\hline
LHC & 11\% & \cite{ATLAS:2023tkt,CMS:2023sdw}\\
High-Luminosity LHC (HL-LHC) & 2.25\% & \cite{Cepeda:2019klc}\\
Compact Linear Collider (CLIC) & 0.69\% & \cite{CLICdp:2018cto}\\
Future Circular Collider (FCC$ee$) &  0.19\% & \cite{Blondel:2021ema}\\
International Linear Collider (ILC) & 0.16\% & \cite{Potter:2022shg}\\
\hline
\end{tabular}
\caption{\label{tab:Hinv} Current and projected limits on the Higgs invisible decay from the LHC, HL-LHC, ILC, CLIC, and FCC-ee \cite{ATLAS:2023tkt,CMS:2023sdw,Cepeda:2019klc,CLICdp:2018cto,Blondel:2021ema,Potter:2022shg}.}
\end{table}

\section{Dark Matter } \label{DM:pheno}
We begin by examining the production mechanism of DM in the early Universe. 
In the case of thermal freeze-out DM, its relic abundance can be determined by solving the Boltzmann equation, which takes the form
\begin{align}
	\frac{dY_X}{dx} &= -\, \frac{s(x)\,\langle \sigma v \rangle_{\rm tot}}{H(x)}
	\left( Y_X^{2} - \left(Y_X^{\rm eq}\right)^{2} \right),
\end{align}
where $x\equiv m_X/T$, $H$ is the Hubble rate and $Y^{\rm eq}$ is the equilibrium comoving density.

Once the co-moving number density is obtained, the present dark matter relic density is related to it via~\cite{Edsjo:1997bg},
\begin{eqnarray}
	\Omega_X h^{2} = 2.755 \times 10^{8}
	\left( \frac{m_X}{\rm GeV} \right) Y_X.
\end{eqnarray}
Considering DM self-annihilation only, the DM annihilation cross section into a pair of charged leptons and neutrinos can be expressed as
\begin{align}
	\sigma(s) &= \sum_{\ell} 
	\frac{\kappa_f g_X^4}{12\pi}\, \beta_f \beta_X 
	\left[
	\frac{s + 2m_f^2}{(s - m_{Z'}^2)^2 + m_{Z'}^2 \Gamma_{Z'}^2}
	\right],
\end{align}
where $\beta_i = \sqrt{1 - m_i^2/s}$, and $f = \mu, \tau, \nu_\mu, \nu_\tau$. 
The total decay width of the $Z'$ boson is given by
\begin{align}
	\Gamma_{Z'} = 
	\frac{\kappa_f g_X^2 m_{Z'}}{12\pi}
	\left( 1 + \frac{2m_f^2}{m_{Z'}^2} \right)
	\sqrt{1 - \frac{4m_f^2}{m_{Z'}^2}} \,
	\Theta\left(m_{Z'} - 2m_f\right),
\end{align}
with $\kappa_f = 1$ for $f = \mu, \tau$ and $\kappa_f = 1/2$ for $f = \nu_\mu, \nu_\tau$. 

Throughout this study, we take $m_{Z'} = 10~\text{MeV}$, and assume that the DM particle is always heavier than $Z'$ boson. 
Consequently, the decay channel $Z' \to X\bar{X}$ is kinematically forbidden. 
When the dark Higgs mechanism is not included, the observed relic abundance of DM can be achieved only if $m_{Z'} \simeq m_X/2$, 
which originates from the smallness of gauge coupling $g_X$ \cite{Foldenauer:2018zrz, Holst:2021lzm,Drees:2021rsg}.
This leads to a strong mass correlation between $Z'$ and DM. 
A possible way to relax this correlation within the dark Higgs framework has been discussed in Refs.~\cite{Baek:2020owl,Baek:2022ozm, Ho:2024cwk}.

In this work, distinctive semi-annihilation channels such as 
\(XX \rightarrow \bar{X} H_1\) and \(XX \rightarrow \bar{X} Z'\) 
are introduced, in addition to the self-annihilation modes 
\(X\bar{X} \rightarrow Z' \phi, Z' h \rightarrow\) SM particles~\cite{Ko:2014nha}. 
The presence of these extra processes enables a wider viable mass range 
for the $Z_3$ complex scalar DM \(X\), thereby alleviating the tight 
mass correlation \(m_X \simeq m_{Z'}/2\) that commonly arises in models 
where only the \(Z'\) mediator is involved and the \(H_1\) field is absent.

\begin{figure}[!t]
	\centering
	\hspace{0.5cm}
	\includegraphics[width=0.95\linewidth]{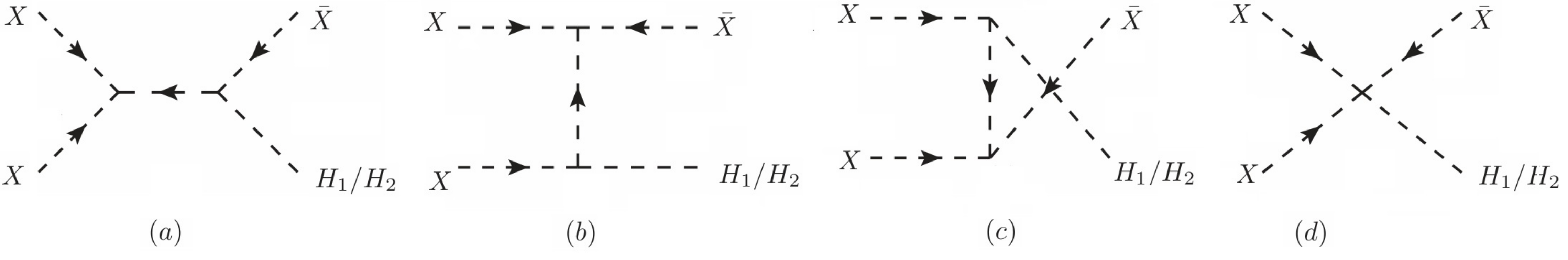}\\
	\includegraphics[width=0.73\linewidth]{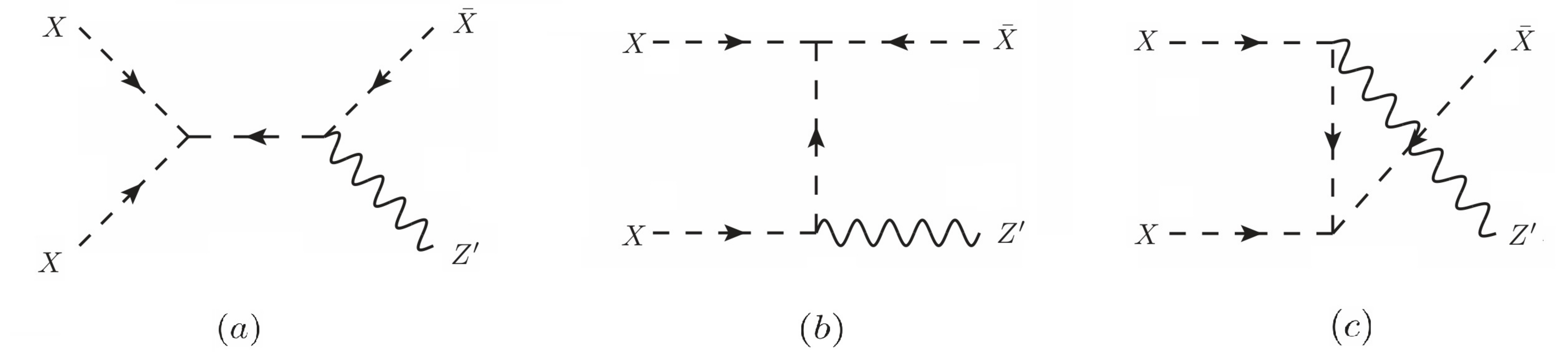}
	\hspace{0.5cm}
	\caption{Feynman diagrams for dark matter semi-annihilation.\label{feynman}}
\end{figure}

We now focus on the scenario in which the DM relic abundance is predominantly governed by semi-annihilation processes in the early Universe  because of the smallness of $g_X$.
In contrast to the conventional self-annihilation, semi-annihilation allows one of the final-state particles to remain a DM particle, accompanied by either a SM field or a new light dark-sector state, thereby naturally accounting for the observed relic density. 
Including such semi-annihilation effects, the Boltzmann equation can be written as
\begin{align}
	\frac{dY_X}{dx} &= 
	-\, \frac{s(x)\,\langle \sigma v\rangle_{X\bar{X}\to {\rm SM}}}{H(x)}
	\left( Y_X^2 - (Y_X^{\rm eq})^2 \right)
	+ \frac{1}{2}\, \frac{s(x)\,\langle \sigma v\rangle_{XX\to \bar{X} Y}}{H(x)}
	\left( Y_X^2 - Y_X Y_X^{\rm eq} \right).
\end{align}
To realize this setup, we take $\lambda_{XH} = \lambda_{X\Phi} = 0$. 
Although DM annihilation into lepton pairs remains kinematically accessible, the corresponding annihilation cross section is typically suppressed, except near the resonance region where $m_X \simeq m_{Z'}/2$. 
The relevant Feynman diagrams are shown in Fig.~\ref{feynman}.

The thermal-averaged semi-annihilation cross sections are described by 
\begin{align}
	\langle \sigma v\rangle_{XX \to \bar{X} Z'} = &\frac{\lambda^2_3}{576\pi m^6_X} \frac{ \left( m^2_X - m^2_{Z'} \right) \left( 9 m^2_X - m^2_{Z'} \right)^3  }{\left( 3m^2_X - m^2_{Z'} \right)^2 }\sqrt{9-\frac{10m^2_{Z'}}{m^2_X} + \frac{m^4_{Z'} }{ m^4_X}  } + \mathcal{O}(v^2), \\
	\langle \sigma v\rangle_{XX \to \bar{X} H_1} =& \frac{\lambda^2_3}{64\pi} \frac{ \left( s_\theta v_H v_\phi \lambda_{XH} \left( 9m^2_X + m^2_{H_1} \right) + c_\theta \left( 3m^2_X \left( m^2_{H_1}+3m^2_X \right) -\lambda_{X\Phi} v^2_\Phi \left( m^2_{H_1}-9m^2_X \right) \right)  \right)^2  }{ m^6_X \left( 3m^2_X - m^2_{H_1} \right)^2 } \nonumber\\
	&\times \sqrt{9-\frac{10m^2_{H_1}}{m^2_X} + \frac{m^4_{H_1} }{ m^4_X}  }  + \mathcal{O}(v^2). 
\end{align}

In the heavy DM mass limit and for $\lambda_{XH}=\lambda_{X\Phi}=0$, the ratio between these semi-annihilations becomes equal due to the Goldstone boson theorem.
The related cross section is given by
\begin{align}
	\langle \sigma v\rangle_{XX \to \bar{X} Z'} = \langle \sigma v\rangle_{XX \to \bar{X} H_1}  =\frac{27 }{64 \pi }\frac{\lambda^2_3}{m^2_X}.
\end{align}
We note that the semi-annihilation cross section is independent of the choice of $v_\phi$in this limit. 
For numerical studies, we make a model file by using FeynRules \cite{Alloul:2013bka}.
Taking into account all of the DM self- and semi-annihilation channels, we obtain the relic density  using micrOMEGAs code \cite{Alguero:2023zol}.
\begin{figure}[!t]
	\centering
	\hspace{0.5cm}
	\includegraphics[width=0.6\linewidth]{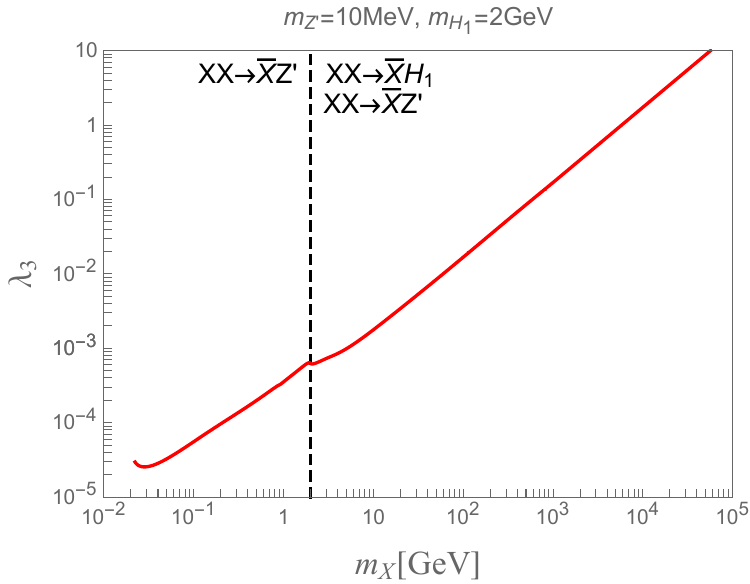}
	\hspace{0.5cm}
	\caption{
		The relic density in the $(m_X,\, \lambda_3)$ plane is shown. We fix $m_{Z'} = 10~\mathrm{MeV}$, $m_{H_1} = 2~\mathrm{GeV}$, and $\sin\theta = 3 \times 10^{-3}$ to account for the Belle~II excess. The impact of varying $\sin\theta$ is negligible due to its small value. The red solid line corresponds to the observed dark matter relic density measured by Planck~\cite{Planck:2018vyg}.  To the left (right) of the dashed vertical line, the relic abundance is dominantly determined by the process $XX \to \bar{X}Z'$ ($XX \to \bar{X}Z',\, \bar{X}H_1$), respectively.
	} \label{semiDM}
\end{figure}

In Fig.~\ref{semiDM}, we show $\Omega_X h^2$ as a function of $\lambda_3$ and $m_X$.
The coupling of $\lambda_3$  controls the strength of semi-annihilation for different choices of $m_X$. 
Given the smallness of the gauge coupling $g_X \leq O(10^{-4})$, the semi-annihilation channel $XX \rightarrow \bar{X} Z'$ provides significant when $m_X < m_{H_1}$.
When $m_X >m_{H_1}$, the process $XX \rightarrow \bar{X} H_1$ is kinematically open.
We see that the dark Higgs can modify DM phenomenology  significantly and the allowed mass range for the complex scalar DM $X$ can significantly deviate from $m_X \sim m_{Z'}/2$ \cite{Baek:2022ozm}.

Although $g_X$ is relatively small, a sizable DM–nucleon elastic scattering cross section can still be obtained compared to Refs.~\cite{Park:2015gdo, Foldenauer:2018zrz,Holst:2021lzm,Drees:2021rsg}. 
This is because a light $Z'$ boson is involved in the $t$-channel of the DM–nucleon scattering process and avoiding the strong mass correlation.
The DM-nucleon elastic scattering is given by
\begin{align}
	\sigma^{X-n}_{\rm el}  &\simeq \frac{\mu^2_n}{\pi}\frac{e^2 g^2_X Z^2 \epsilon^2_A }{ A^2 m^4_{Z'} },
\end{align}
where $\mu_i$ is the reduced mass of DM and particle $i$, $A$ and $Z$ are the number of proton and the nucleus, respectively.
When DM mass is above 10GeV, the most stringent bound on DM-nucleon elastic scattering is from 4.2 Tonne-Years  of Exposure of the LUX-ZEPLIN experiment \cite{LZ:2024zvo}.
Similarly, the DM–electron elastic scattering cross section is given by
\begin{align}
	\sigma^{X-e}_{\rm el}  &\simeq \frac{\mu^2_e}{\pi}\frac{e^2 g^2_X  \epsilon^2_A }{ m^4_{Z'} } = \frac{m_e}{\pi m_X }\frac{e^2 g^2_X  \epsilon^2_A }{ m^4_{Z'} },
\end{align}
where we have used $\mu_e \to m_e/m_X$ when DM mass is much heavier than the electron mass.
The current most stringent bound on DM-electron elastic scattering comes from DAMIC-M \cite{DAMIC-M:2025luv}.
We adpoted DM-electron bound from DAMIC-M for $m_X < 1$GeV.
The dark matter elastic scattering bounds from LZ (DAMIC) are 
$\sigma_{\rm el} \sim 10^{-47}\,(10^{-38})~{\rm cm}^2$ 
for dark matter masses of $36~{\rm GeV}$ ($10~{\rm MeV}$), respectively.

\begin{figure}[!t]
	\centering
	\hspace{0.5cm}
	\includegraphics[width=0.6\linewidth]{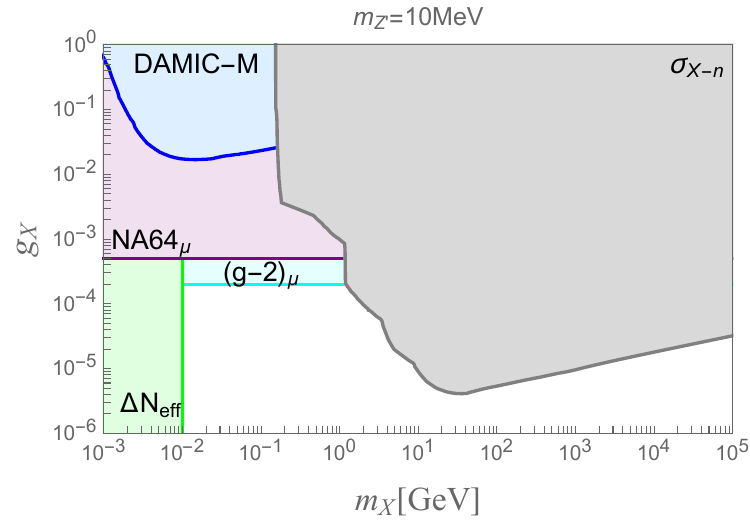}
	\hspace{0.5cm}
	\caption{
		Allowed $(m_X,~g_X)$ space when $m_{Z'} = 10~\mathrm{MeV}$. The blue-shaded region is excluded by the DAMIC-M result~\cite{DAMIC-M:2025luv}, while the green area is ruled out by the bound $\Delta N_{\rm eff}$. The purple region is excluded by the NA64~\cite{NA64:2024klw}. The light-cyan band is disfavored by the latest muon $(g-2)$ measurement at $\Delta a_\mu > 10^{-10}$. Finally, the gray-shaded area is constrained by the current dark matter direct detection limits.
	} \label{mZpVsgX}
\end{figure}

In Fig.~\ref{mZpVsgX}, we depict the allowed parameter space. 
Here we take $m_{Z'}=10$MeV, the allowed value of $g_X$ can sustainably relax the Hubble tension \cite{Kamada:2015era, Escudero:2019gzq}. 
The green shaded area is excluded by $\Delta N_{\rm eff}$ from the Planck observation.
The purple region is excluded by ${\rm NA64}_\mu$ experiment data \cite{NA64:2024klw}.
Taking into account new SM $(g-2)_\mu$ prediction, the cyan area is ruled out \cite{Aliberti:2025beg}.
The blue shaded region is excluded by the DM-electron scattering bound from DAMIC-M \cite{DAMIC-M:2025luv}.
The gray region is ruled out by the DM direct detection bound on DM-nucleon scattering process by LUX-ZEPLIN \cite{LZ:2024zvo}. 
Notice that the viable dark matter mass range strongly depends on the gauge coupling $g_X$. 
For $g_X = 10^{-4}$, the dark matter mass is constrained to be below a few GeV. 
As the coupling decreases to $g_X = 10^{-5}$, the DSM mass up to roughly $\sim 10~\mathrm{GeV}$ is allowed. 
For even smaller couplings, such as $g_X = {\rm a~few} \times 10^{-6}$, the DM mass can be as large as the unitarity limit, $\sim 100~\mathrm{TeV}$.
Finally, we briefly discuss the constraint arising from DM annihilation bound at the 
CMB epoch. 
Dark matter annihilation during the recombination era, $T\sim {\rm eV}$, can inject energy to ionizing particles, 
which in turn modifies the ionization history of the Universe and affects the temperature and polarization anisotropies of the CMB~\cite{Slatyer:2015jla}. 
These effects are tightly constrained by the most recent Planck satellite observations, 
which restrict the total amount of energy deposition in that epoch. 
The constraint is conventionally parametrized as~\cite{Planck:2018vyg}
\begin{align}
	p_{\rm ann} \equiv f_{\rm eff}\,\frac{\langle\sigma v\rangle}{m_X},
\end{align}
where $f_{\rm eff}$ denotes the effective efficiency factor that quantifies the fraction of the annihilation energy deposited into the intergalactic medium. 
According to the Planck 2018 results, the 95\%~C.L. upper bound is~\cite{Planck:2018vyg}
\begin{align}
	p_{\rm ann} \lesssim 3.5\times10^{-28}~\mathrm{cm^3\,s^{-1}\,GeV^{-1}}.
\end{align}
This constraint can rule out light dark matter scenarios with masses below $\sim 10$ GeV, where DM annihilates into electromagnetically charged final states via the $s-$wave processes.
In the present framework, however, this severe constraint is naturally avoided, since the dark photon $Z'$ and dark Higgs $H_1$ produced via semi-annihilation predominantly decay into neutrinos.  

Previous studies~\cite{Escudero:2018mvt,Chu:2023jyb} have shown that light dark matter annihilating into neutrinos can alter the effective number of relativistic species, $N_{\rm eff}$. In particular, Ref.~\cite{Chu:2023jyb} showed that a complex scalar dark matter with a mass below $8.2\,{\rm MeV}$ is disfavored due to its impact on the CMB. In this work, we focus on the regime $m_X \gg 10\,{\rm MeV}$, where such constraints are no longer relevant, and thus the scenario naturally avoids the bound from $N_{\rm eff}$.
\section{Belle II Excess  }\label{Belle2}
After the recent announcement from the Belle~II collaboration, the measured branching fraction ${\rm Br}(B^+ \to K^+ \nu \bar{\nu})$ shows a $2.7\sigma$ deviation from the SM expectation~\cite{Belle-II:2023esi}.
The excess observed by Belle~II may hint at the presence of new physics beyond the Standard Model. 
If such new physics contributes to the $b \to s \nu \bar{\nu}$ transition, appearing as missing energy $\slashed{E}$ 
in the final state of the $B^+ \to K^+$ decay. 
The corresponding branching fraction induced by new physics effects can be written as
\begin{eqnarray}
	{\rm Br}\!\left(B^+ \to K^+ \slashed{E}\right)_{\rm NP} 
	= (1.8 \pm 0.7) \times 10^{-5}.
\end{eqnarray}
Ref.~\cite{Altmannshofer:2023hkn} emphasized that the Belle~II analysis provides valuable information on the reconstructed invariant-mass spectrum $q^2_{\rm rec}$, 
which exhibits a localized enhancement around $q^2_{\rm rec} \simeq 4~{\rm GeV}^2$. 
Such a feature can be explained by introducing a new particle with a mass of $\sim 2~{\rm GeV}$, 
provided that its couplings to the SM fermions are sufficiently suppressed or absent. 
Including the null result from BaBar data, their combined global analysis yields a branching ratio 
\begin{align}
	{\rm Br}(B^+ \to K^+ +\chi) = (5.1 \pm 2.1) \times 10^{-6},
\end{align}
where $\chi$ denotes the new hidden particle. 
In the case of the two-body decay scenario, the statistical significance decreases to $2.4\sigma$. 
Therefore, distinct branching ratios are expected between the two-body and three-body decay interpretations.
We consider two different branching ratios corresponding to two-body and three-body decays.
In particular, Ref.~\cite{Ho:2024cwk} demonstrated that both the observed dark matter relic density and the Belle~II excess can be simultaneously accommodated within a unified framework for the first time.

First, let us focus on the two-body decay scenario. 
In this case, the $B^+$ meson decays into a kaon and a dark Higgs boson, $B^+ \to K^+ H_1$. 
The dark Higgs mass is determined to be $m_{H_1} = 2~\mathrm{GeV}$ from the observed $q^2$ spectrum \cite{Altmannshofer:2023hkn}.
Such a decay is kinematically allowed when $m_{B^+} - m_{K^+} > m_{H_1}$. 
Within our framework, the decay width for this two-body decay can be expressed as
\begin{align}
	\Gamma(B^+ \to K^+ H_1) \simeq 
	\frac{|\kappa_{cb}|^2 \sin^2\!\theta}{64\pi m_{B^+}^3}
	\left[ f_0(m_{H_1}^2) \right]^2
	\left( \frac{m_{B^+}^2 - m_{K^+}^2}{m_b - m_s} \right)^2
	\sqrt{\lambda(m_{B^+}^2, m_{K^+}^2, m_{H_1}^2)},
\end{align}
where $\kappa_{cb} \simeq 6.7 \times 10^{-6}$ represents the one-loop induced vertex factor obtained after integrating out the top quark and $W$ boson. 
The function $f_0(q^2)$, with $q^2 = (p_B - p_K)^2$, denotes the $B^- \to K^-$ transition form factor~\cite{Parrott:2022rgu}, 
and $\lambda(a,b,c) \equiv a^2 + b^2 + c^2 - 2(ab + bc + ac)$ is the usual K\"all\'en function.
%

Let us now turn to the three-body decay scenario. 
In this case, the $B^+$ meson decays via $B^+ \to K^+ H^{(*)}_1 \to K^+ Z' Z'$ or $K^+ X\bar{X}$. 
Since the dark Higgs does not couple directly to the dark matter pair at tree level, 
it cannot decay into $X\bar{X}$ in the absence of the relevant portal couplings.
The decay width for the process $B^+ \to K^+ Z' Z'$ is given by
\begin{align}
	\Gamma\!\left(B^+ \to K^+ Z'Z' \right) 
	=&~ \frac{ |\kappa_{cb}|^2  \left( 2 g_X Q_\Phi m_{Z'} s_\theta c_\theta \right)^2 }{512 \pi^3 m_B^3}
	\int dm_{23}^2 \int dm_{12}^2 
	\left( 3+\frac{ m_{12}^2 (m_{12}^2 - 4m_{Z'}^2) }{4 m_{Z'}^4} \right)
	\nonumber\\
	&\times 
	\left( \frac{1}{m_{12}^2 - m_{H_1}^2 } 
	- \frac{1}{m_{12}^2 - m_{H_2}^2 } \right)^{\!2} 
	\left[f_0(m_{12}^2)\right]^2
	\left( \frac{m_{B^+}^2 - m_{K^+}^2}{m_b - m_s} \right)^{\!2}.
\end{align}
When the portal couplings $\lambda_{HX}$ and $\lambda_{X\Phi}$ are set to zero, 
the three-body decay channel $B^+ \to K^+ H^{(*)}_1 \to K^+ Z' Z'$ is already excluded 
by the upper limit on the invisible decay width of the Standard Model Higgs. 
For non-zero values of $\lambda_{HX}$ and $\lambda_{X\Phi}$, however, 
a viable parameter space might remain~\cite{Ho:2024cwk}.

\begin{figure}[!t]
	\centering
	\hspace{0.5cm}
	\includegraphics[width=0.47\linewidth]{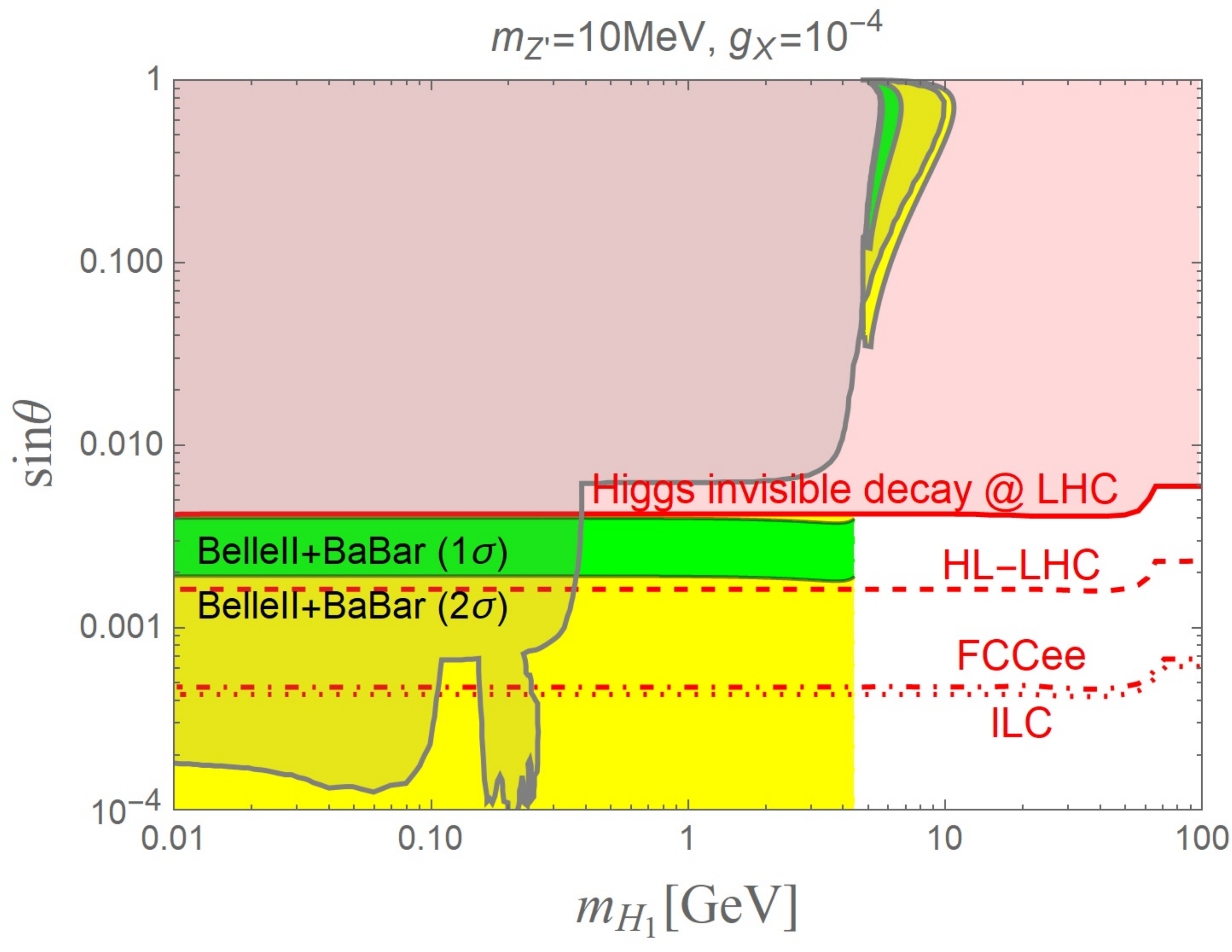}
	\includegraphics[width=0.47\linewidth]{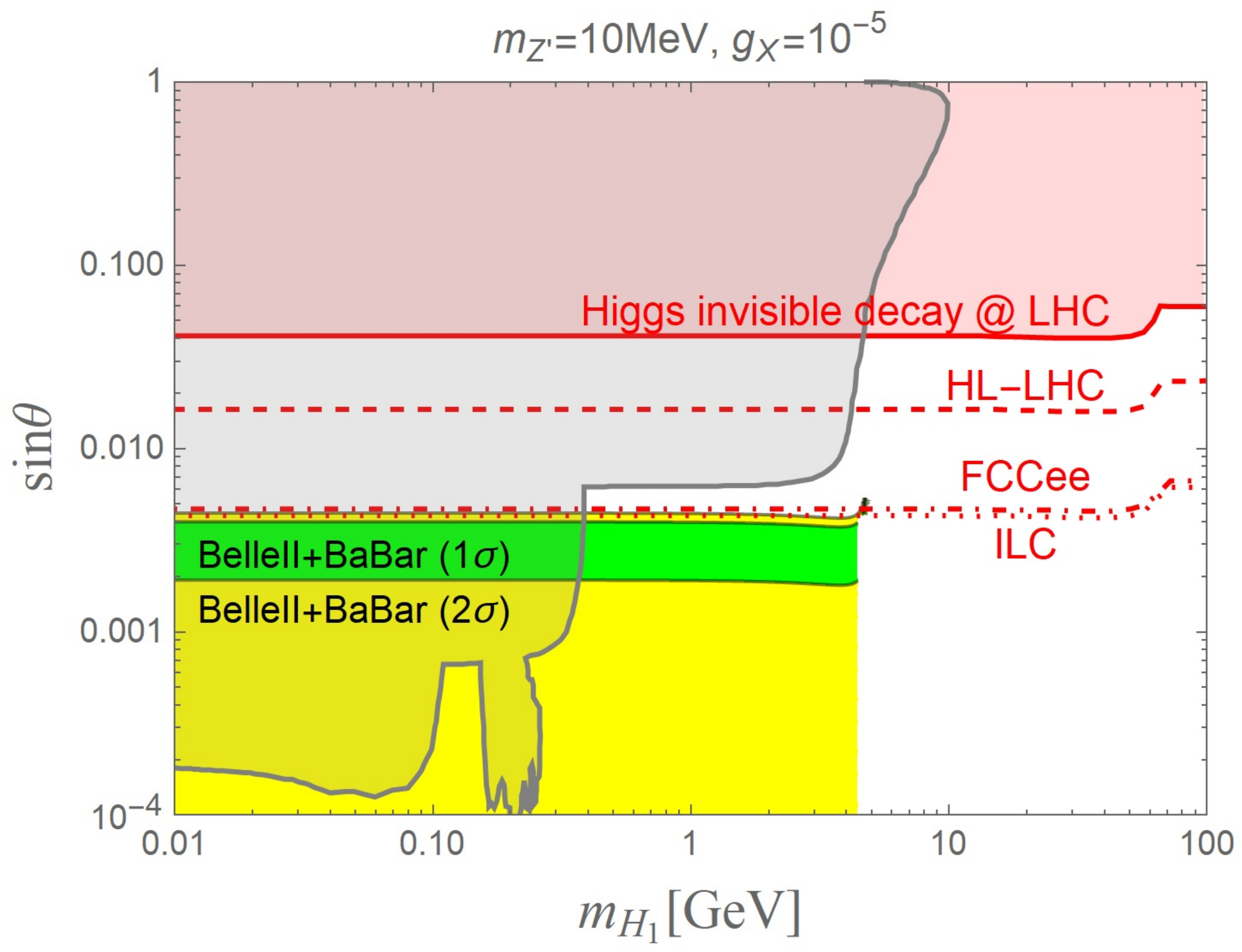}
	\hspace{0.5cm}
	\caption{
		The regions allowed at the $1\sigma$ and $2\sigma$ CL by the Belle~II excess correspond to the areas enclosed within the green (inner) and yellow (outer) shaded bands, respectively. 
		We adopt $g_X = 10^{-4}$ for the left panel and $g_X = 10^{-5}$ for the right panel. 
		The red-shaded region is excluded by the constraint from the invisible decay width of the SM Higgs boson~\cite{ATLAS:2023tkt,CMS:2023sdw}. 
		For 3-body decay case, the preferred region is already ruled out by the Higgs invisible decay bound. 
		However, the parameter space can be partially reopened when the couplings $\lambda_{X\Phi}$ and/or $\lambda_{H\Phi}$ are nonzero. 
		The gray-shaded area denotes the parameter region excluded by $K^+ \to \pi^+ + {\rm inv.}$, $K^0_L \to \pi^0 \nu\bar{\nu}$, and $B^0 \to K^{*0} \nu\bar{\nu}$~\cite{NA62:2021zjw, KOTO:2020prk,Belle:2017oht}. \label{BelleII}
	} 
\end{figure}
In Fig.~\ref{BelleII}, we show the preferred region of BelleII excess at the $1\sigma$ and $2\sigma$ confidence levels, shown as the green (inner) and yellow (outer) shaded regions, respectively. 
The left and right panels correspond to $g_X = 10^{-4}$ and $g_X = 10^{-5}$. 
The red-shaded region is excluded by constraints from the invisible decay width of the Standard Model Higgs boson~\cite{ATLAS:2023tkt,CMS:2023sdw}. 
This bound becomes weaker as the coupling $g_X$ decreases. 
In the three-body decay scenario, the preferred region is already ruled out by the Higgs invisible decay limit.
A portion of the parameter space might be recovered when the couplings $\lambda_{X\Phi}$ and/or $\lambda_{H\Phi}$ are nonzero. 
Finally, the gray-shaded region represents the parameter space excluded by rare meson decays, including $K^+ \to \pi^+ +{\rm inv.}$, $K^0_L \to \pi^0 \nu \bar{\nu}$, and $B^0 \to K^{*0} \nu \bar{\nu}$~\cite{NA62:2021zjw, KOTO:2020prk,Belle:2017oht}.

\section{Conclusions  }\label{con}
Following the announcement of the Belle~II excess, a few studies have attempted to simultaneously account for both the observed DM relic density and the ${\rm Br}(B^+ \to K^+ \nu\bar{\nu})$ excess from Belle~II ~\cite{Ho:2024cwk, He:2024iju,Calibbi:2025rpx, Ding:2025eqq}.  
In most cases, explaining the Belle~II excess with light dark matter leads to an 
excessively large relic abundance due to the requirement of the small coupling. 
In order to reproduce the observed relic density, it is generally necessary to introduce additional dark matter annihilation channels or allow the dark matter to decay.

In search of a viable solution, we consider a scenario in which the $U(1)_{L_\mu - L_\tau}$ symmetry is broken down to local $Z_3$ symmetry. 
Within this framework, two semi-annihilation channels for dark matter are available: $XX \to \bar{X} Z'$ and $XX \to \bar{X} H_1$. 
By taking a light $Z'$ mass to address the Hubble tension and $m_{H_1} = 2~\mathrm{GeV}$ to fit the Belle~II excess, both the Belle~II excess and the correct dark matter relic density can be explained simultaneously. 
The constraint from the SM Higgs invisible decay depends on the choice of the coupling $g_X$. 
For $g_X \lesssim \text{a few} \times 10^{-5}$, even future bounds from the ILC are not expected to exclude the region relevant for the Belle~II excess.

Near galactic center (GC), dark matter can undergo semi-annihilation processes, producing light dark sector particles such as $Z'$ and $H_1$, which subsequently decay into neutrinos. 
Currently, the IceCube experiment places an upper bound on dark matter annihilation directly into a pair of neutrinos at the level of $\langle \sigma v \rangle_{X\bar{X} \to \nu\bar{\nu}} \sim 10^{-24}~\mathrm{cm^3/s}$ \cite{IceCube:2023ies}. 
Compared to DM self-annihilation into a pair of neutrino, the morphology of the produced neutrinos can be different.
A dedicated search for neutrinos originating from DM semi-annihilation in the GC would be an interesting possibility, and we leave a detailed study of this scenario for future work.

\acknowledgments
The work is supported in part by Basic Science Research Program through the National Research Foundation of Korea (NRF) funded by the Ministry of Education, Science and Technology (RS-2024-00341419) [JK], and (RS-2025-24803289) [PK], 
and by KIAS Individual Grant No. PG021403.
This research is also supported  in part by the Excellence Cluster ORIGINS which is funded by the Deutsche Forschungsgemeinschaft (DFG, German Research Foundation) under Germany's Excellence Strategy–EXC-2094-390783311 [JK,PK].

\bibliographystyle{JHEP}
\bibliography{literature}

\end{document}